\documentclass[12pt]{article}
\usepackage{rotating}
\usepackage{epsfig,amssymb}
\input{epsf}

\def\laq{\raise 0.4 ex \hbox{$<$}\kern -0.8 em\lower 0.62 ex\hbox{$\sim$}}
\def\gaq{\raise 0.4 ex \hbox{$>$}\kern -0.7 em\lower 0.62 ex\hbox{$\sim$}}

\def\beq{\begin{equation}}
\def\eeq{\end{equation}}
\def\beqa{\begin{eqnarray}}
\def\eeqa{\end{eqnarray}}

\def\to{\rightarrow}

\def\to{\rightarrow}

\def\Ps{\Psi}

 \def\frac#1#2{{\textstyle{{#1}\over {#2}}}}
 \def\lsim{\mathrel{\rlap{\lower4pt\hbox{\hskip1pt$\sim$}}
    \raise1pt\hbox{$<$}}} \def\gsim{\mathrel{\rlap{\lower4pt\hbox{\hskip1pt$\sim$}}
    \raise1pt\hbox{$>$}}}
\def\sqr#1#2{{\vcenter{\vbox{\hrule height.#2pt
         \hbox{\vrule width.#2pt height#1pt \kern#1pt
         \vrule width.#2pt}
         \hrule height.#2pt}}}}

\def\gappeq{\mathrel{\rlap {\raise.5ex\hbox{$>$}} {\lower.5ex\hbox{$\sim$}}}}

\def\lappeq{\mathrel{\rlap{\raise.5ex\hbox{$<$}}
{\lower.5ex\hbox{$\sim$}}}}

\begin{document}
\pagestyle{plain}

\begin{flushright}
September 2022
\end{flushright}
\vspace{5mm}

\begin{center}

{\Large\bf From the Stochastic Weather to a Putative Chaotic Earth System\footnote{Based on talk delivered at the Green Marble 2022, ``International Meeting on Anthropocene Studies and Ecocriticism: Only One Earth", June 30th - July 2nd, 2022, Porto, Portugal.}}

\vspace*{0.2cm}

Orfeu Bertolami$^{1,2}$  \\
\vspace*{0.5cm}
{$^1$Departamento de F\'\i sica e Astronomia, Faculdade de Ci\^encias, Universidade do Porto\\
Rua do Campo Alegre s/n, 4169-007 Porto, Portugal}\\
\vspace*{0.2cm}
{$^2$Centro de F\'\i sica das Universidades do Minho e do Porto, Faculdade de Ci\^encias, Universidade do Porto\\
Rua do Campo Alegre s/n, 4169-007 Porto, Portugal}\\

\vspace*{0.2cm}
\end{center}

\begin{abstract}

\noindent
In this brief report we discuss how continuous changes on the physical parameters that determine the weather conditions may lead to long term climate variability. 
This variability of the weather patterns  are a response to continuous random short period weather excitations that are imprinted in the 
ocean-atmosphere-cryosphere-land system, the Earth System. Given that Earth System is, in the Anthropocene, dominated  by the human action, it  
responds to the intensity and the rate of change of the humankind activities. Thus, we argue, in the context of a specific model of the Earth System, 
that this rate of change may admit a chaotic-type behaviour.

\end{abstract}

\section{Introduction}

Weather conditions depend on physical parameters that change continuously.  These changes may lead to the misconception that the effects on the patterns of the weather are in fact just a 
result of these very changes. 
However, it was shown in the pioneering work of Hasselmann \cite{Hasselmann} that long term climate variability can be accounted as the response to 
continuous random short period weather excitations. This response is the result of the integration of the rapidly varying weather system, essentially 
the atmosphere, and the slowly responding climate system (the ocean, cryosphere, land vegetation, etc) on the coupled ocean-atmosphere-cryosphere-land system. 

Indeed,  in dynamical weather models where only the average transport effects of
the rapidly varying weather components are parameterised in the climate system, the resulting 
equations are deterministic, and climate variability arise only through variable external conditions. 

The main new feature brought by stochastic climate models approach analysed by Hasselmann 
is that the non-averaged weather components are also retained and 
they appear formally as random forcing terms. Thus, the climate system, acts as an integrator
of the short-period excitations and exhibits the same ``random-walk" response, a forced one though. 

In fact, the problem was shown to be analogous to the Brownian motion problem, where large particles 
interact with an ensemble of  much smaller particles. The evolution of the climate probability
distribution is described by a Fokker-Planck equation and the resulting model predicts a ``red" variance
spectra in agreement with the observations, namely an increase in climate variability. 
In other words, the random weather variability can give rise to systematic changes of the patterns of the weather (humidity, droughts, heat waves, catastrophic 
events, etc) once a forcing with a well defined feature.

The theoretical work of Hasselmann gives support to the increasingly sophisticate numerical models developed by Manabe and collaborators, who since 1965 have 
shown that a clearly climate variability would arise from an increase on the concentration of greenhouse gases in the atmosphere  with the net effect 
of raising the global temperature. (see Ref. \cite{Manabe} for a detailed account). 

Meanwhile, evidence about the complexity of the coupling of the various components of the Earth System was strengthened and it became consensual 
that the main driver of the  Earth System (ES) are no longer the natural forces of astronomical or geological nature, but instead the human activities. 

Indeed, the understanding of the processes that shape the ES involve interactions and feedbacks through which material and energy fluxes 
among the Earth's sub-components namely the atmosphere, biosphere, cryosphere, hydrosphere, magnetosphere,  
upper iithosphere, as well as the impact of the activities of the human societies. Of course, the spatial configuration and the time 
evolution of the ES sub-systems, and their variability and stability must be seen in a holistic way. 

Notice that the time scale issue is crucial in setting the spatial impact of a given local event: at a short time scale its implications are spatially restricted; 
however, on a longer term, the resulting impact may propagate to the entire ES. Thus, in oder to properly describe the ES it is important that its description exhibits some 
general features:

\vspace{0.2cm}

\noindent 1) Its variability across space and time has been profoundly affected by the impact of recent human activities since 1950s or so, 
which is seriously compromising the stability of the ES at the Holocene (the last 11700 years);

\vspace{0.2cm}

\noindent 2) The effect of the biological processes are, contrary to what was initially envisaged, as relevant as the ones of astronomical and geological nature;

\vspace{0.2cm}

\noindent 3) High connectivity among all of its components; 

\vspace{0.2cm}

\noindent 4) Its behaviour is highly non-linear, meaning that small changes due to a forcing factor can push the system across a new threshold. 
In fact, we are witnessing the crossing of tipping points by a quite visible climate change/global warming, driven 
by the systematic increase in the concentration of greenhouse gases in the atmosphere. 

The idea of a highly connected planet makes part of the cosmogony of many ancient civilisations (see for instance, \cite{Bertolami2006}). A holistic view of the natural processes on 
Earth was defended by the 19th century naturalist and geographer, Alexander von Humbold (1769 -- 1859), and in the 20th century, more specifically 
in what concerned the biosphere as a geological force, by the Ukrainian mineralogist and geochemist Vladinir Vernadsky (1863 -- 1945). 
However, it was only in mid 1960s that the crucial biosphere feedback on the ES was conjectured by James Lovelock (1919 -- 2022) through his Gaia Hypothesis:  living organisms and 
their inorganic surroundings form a synergistic and self-regulating complex system that maintains the conditions for life on Earth \cite{Lovelock}. Despite the initial scepticism, a decade later the 
ideias of Lovelock were fully vindicated by the gathering of the various NASA's programmes to observe Earth from space. The resulting report \cite{NASA} is 
regarded as the inaugural document of the ES science.  For a full account of the history line of the ES science see, for instance, Ref. \cite{Perspective}.

Subsequent efforts aimed to deepen the knowledge of the interaction mechanisms between the ES components and to device models. 
Naturally, the main purpose of the models is to provide a suitable description of the global implications of causes that affect some components of the ES 
and, above all, to provide scenarios for its evolution, considering, in particular, that human activities are driving the ES farther away from the stable Holocene conditions. 

In the next section we shall describe a ES model developed here in Porto. It allows for making predictions based on the evolution of the human activities, expressed 
in terms of the so-called Planetary Boundaries \cite{Steffen1,Rockstrom,Steffen2} (see below). In section 3, we shall discuss how, 
through some fairly broad assumptions, one might envisage that the ES can become 
unpredictable and behave in a chaotic way. In section 4, we shall present a broad outline of the features arising from our description. 

\section{The Anthropocene Equation}\label{Sec:II}

Given the complexity of the ES and the elaborate web of interactions among its components, models to describe the behaviour of the ES most often
materialise themselves through complex computer simulations, which allow for tracking the evolution of the salient properties of the ES.  
In what follows we shall describe a simple analytical approach to capture the main features of the ES based on quite broad thermodynamic considerations that are 
ultimately at the core of all the processes that take place in the ES. The fundamental assumption of our model is that the major transformations that took place throughout Earth's 
weather history were, in fact, phase transitions likewise the ones that change the state of the materials, i.e. solid-liquid-gas transformations. The general theory of these  
transitions was proposed by Lev Landau in 1937 and later extended to encompass metastable states by Vitaly Ginzburg in 1960. 
This formalism allows one to write from first principles the free energy function, $F$, the quantity that accounts for the overall energy transformations 
due to the variation of the internal energy, dissipation due to the growth of entropy and the work that can be performed by the system. Of course, the free energy must 
depend on the dynamical variables that affect the system and on its temperature. 

An implementation of this approach to the ES was performed in Ref. \cite{Bertolami2018} and the relevant variables to consider are: 
astronomical parameters of the Earth's orbit around the Sun, our main source of energy; geological and internal parameters; the temperature, $\Psi$; and the human activities, $H$. 
Denoting astronomical and geological parameters collectively as $\eta$, then in general the free energy can be expressed as $F=F(\eta,\Psi,H)$. In order to understand 
the time evolution of ES, one needs to consider the time derivative of $F$, which yields a function, $f(\eta,\Psi,H)$. 
The so-called Anthropocene equation arises after observing that the natural drivers of the ES, $\eta$, have remained essentially unchanged during the Holocene. Thus, we get:
\begin{equation}
{dF \over dt} = f(\eta,\Psi,H)  = lim_{\eta \to 0} f(\eta,\Psi,H) = f(\Psi, H) .
 \label{Anthropocene}
\end{equation}

The predictive power of the Landau-Ginzburg model and hence of our Anthropocene equation derives from the assumption that the free energy 
depends analytically on an order parameter and its spatial derivatives. Our hypothesis is that the order parameter should be the temperature, 
more specifically we choose that $\Psi = (T - \langle T_{\rm H}\rangle)/ \langle T_{\rm H}\rangle$, $ \langle T_{\rm H}\rangle$ 
is the average temperature at the Holocene. Hence, after dropping the derivative terms, one gets \cite{Bertolami2018}:
\begin{equation}
F = F_0 + a(\eta) \Psi^2 + b(\eta)\Ps^4 - H\Psi ,
 \label{F}
\end{equation}
where $F_0$ is an arbitrary constant value, coefficients $a(\eta)$ and $b(\eta)$ can, in principle, be determined from the data (see e.g. Ref. \cite{Bertolami2018}). 
Notice that the last linear term, due to the human activities, is clearly a destabilising factor. The behaviour of the free energy can be better understood through Figure 1.

\begin{figure}
\centering
\includegraphics[width=\columnwidth]{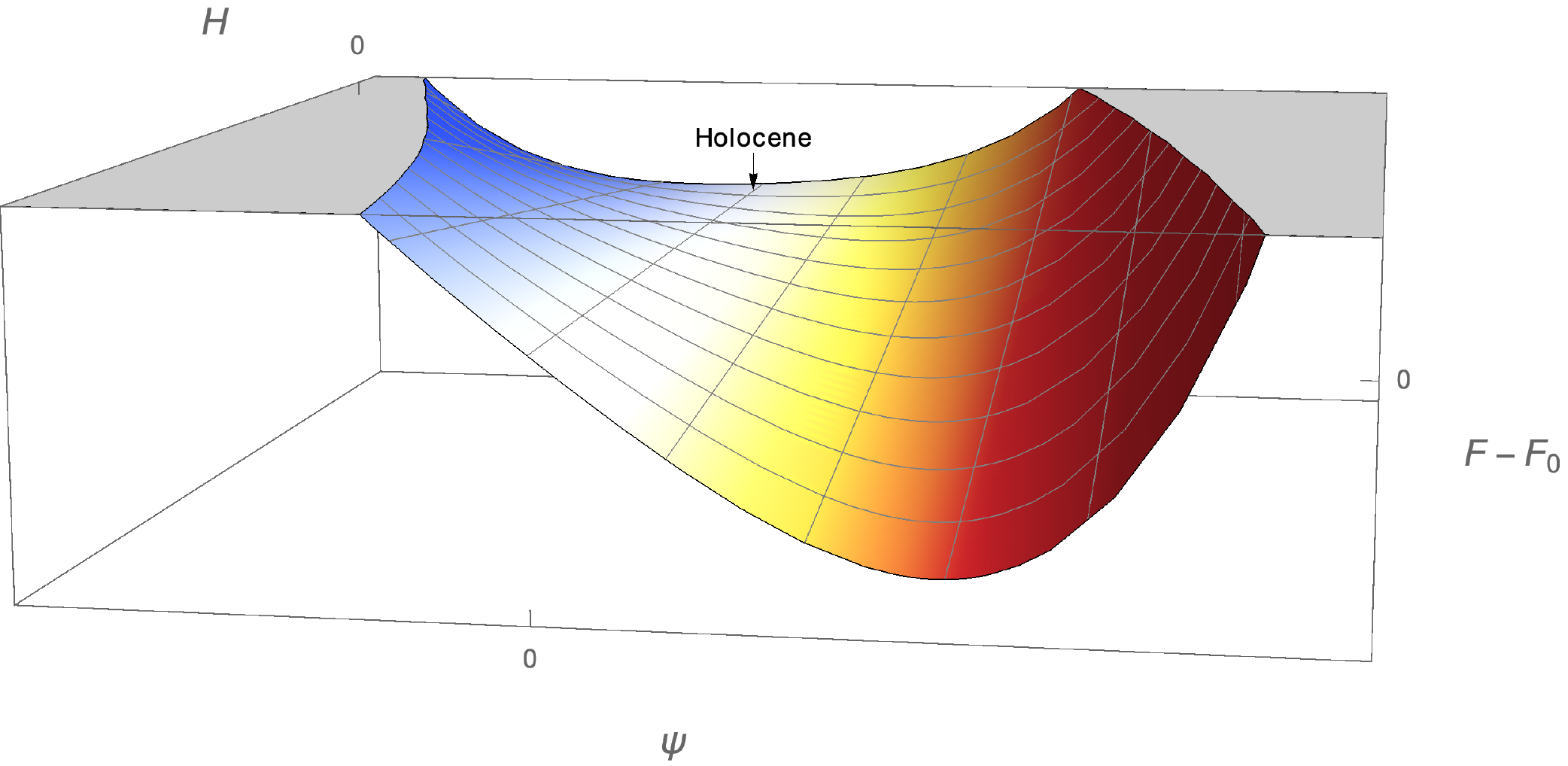}
\caption{Stability landscape of the ES in terms of $\Psi$ and $H$. For $H=0$, the Holocene is clearly a stable state. As $H$ increases, the new minimum (indicated in red) is deeper 
and hotter  \cite{Bertolami2018}.}
\label{fig:stability_landsacpe}
\end{figure}

The proposed model allows for the understanding of the qualitative features of the ES, most particularly the trajectories of the ES in the Anthropocene. 
Indeed, through the assumption that the human activities grow linearly, namely that \cite{Bertolami2019}:
\begin{equation}
H(t) =H_0 t  ,
 \label{Anthropocenelinear}
\end{equation}
where $H_0$ is an arbitrary constant, it was possible to show that the ES is heading, irrespective of the initial conditons at the Holocene, 
towards a new minimum whose temperature is necessarily higher than the Holecene one. 
Hence, human activities lead the ES to an inevitable global warming. This conclusion is 
consistent with the observations and with analysis of Ref.\,\cite{Steffen:2018} concerning the behaviour of the ES in the Anthropocene. 

In order to realistically access the behaviour of the ES in terms of the specific features of its characterisation, one needs to parametrise the impact of the human activities 
on the properties of the ES. This is achieved through the so-called Planetary Boundaries  \cite{Steffen1,Rockstrom,Steffen2}, that is, 
the present values of the minimal set of nine parameters, $h_i$, $i=1, 2, ..., 9$, with respect to their values at the Holocene. These parameters capture the 
complexity of the ES, and comprise the following properties: climate change (CO2 concentration), biodiversity loss, nitrogen and
phosphorus cycles, ocean acidification, land use, fresh water use, ozone depletion, atmospheric
aerosols and chemical pollution. As discussed in Refs. \cite{Steffen1,Rockstrom,Steffen2}, Holocene-like values have kept the ES in a 
favourable operational state, a state where a stable average temperature was reached via feedbacks and forcing driven 
by natural causes and the main regulatory ecosystems of the ES.

In terms of the Planetary Boundaries, we have proposed to specify the function $H(t)$ as follows \cite{Bertolami2019}:  
\begin{equation}
	H = \sum_{i=1}^{9} h_i + \sum_{i,j=1}^{9} g_{ij} h_i h_j + \sum_{i,j,k=1}^{9} \alpha_{ijk} h_i h_j h_k + \ldots,
	\label{eqn:human_action}
\end{equation}
where the second and third set of terms concern the interaction among the various effects of the human action on the Planetary 
Boundaries. Of course, higher order interactions terms can be considered. For most of the purposes, such as for accountancy 
of the human  impact, up to second order and, in fact, a subset of parameters suffices (see e.g. Ref. \cite{Bertolami2020}). It is physically reasonable  
to assume that the $9 \times 9$ matrix, $[g_{ij}]$ is symmetric, $g_{ij} = g_{ji}$, and non-degenerate, $\det[g_{ij}] \neq 0$.

Notice that the evolution of the temperature can be estimated as the equilibrium state evolves as $\langle \psi \rangle \sim H^{1 \over 3}$ \cite{Bertolami2018}. 
Thus, since the effect of the human activities do lead to an increase of $1^{\circ} C$ since the beginning of the Anthropocene \cite{Jones:2004}, 
about 70 years ago, then one should expect in 2050 a temperature increase of $\sqrt[3]{2} = 1.26^{\circ} C$, assuming that the growth of $H$ remains linear \cite{Bertolami2019}.

In what follows we shall consider the implications of replacing the assumption Eq. (\ref{Anthropocenelinear}) by the so-called logistic map. 

\section{Putative Chaotic Trajectories of the Earth System in the Anthropocene}\label{Sec:III}

Before addressing the complex behaviour of the ES that may arise from a more realistic assumption about the evolution of the human activities, let us discuss some of the 
previous knowledge about of the choice we will consider. In fact, one of the first manifestations of the chaotic behaviour were encountered precisely in the 
relatively simple weather model setup by Edward Lorentz in the early 1960s. Indeed, to his surprise, he found that the outcome of this weather 
imulations were extremely sensitive to the initial conditions, 
that is, minute modifications of decimal places digits in the input data of his computer model would lead that the weather patterns they yield would grew farther and farther apart.

This behaviour was quite surprising as it was believed that physics, prior to quantum mechanics, was deterministic, 
meaning that the outcome of the physical equations would be unique and the future would be completely determined by the solution of the 
physical equations and the knowledge of the initial conditions (values of coordinates and the respective velocities in phenomena described by Newtonian physics). 
But soon, Edward Lorentz found out that the odd behaviour he have encountered in his models were actually a feature shared by much simpler models 
involving non-linear evolution equations. Later, this behaviour was referred to as ``Butterfly Effect": the air stirred by a butterfly today in some place, can 
give origin to a storm thousand of kilometers away in  the following month. A decade later, similar surprising features were found by Robert May, 
a mathematician working on ecological population models, in the context of the so-called logistic difference equations (see below). 

The so-called logistic equation was first introduced by the Belgium mathematician, Pierre François Verhulst between 1838 and 1847, as a an improvement of 
model of exponential population growth suggested by Thomas Malthus in 1798 that was based on the hypothesis that the population growth 
at a given time, $t$, would depend only on the number of individuals, $N(t)$, in the population, that is: 
\begin{equation}
{dN(t) \over dt} = r N(t),
 \label{Malthus}
\end{equation}
where the parameter, $r$, is the relative rate of growth of the population. 

On its hand, the logistic solution of Verhulst arises from the differential equation: 
\begin{equation}
{dN(t) \over dt} = r N(t) [1-N(t)].
 \label{Verhulst}
\end{equation}

Actually, Robert May \cite{May} considered the discrete version of Eq. (\ref{Verhulst}), the so-called logistic difference equation or simply Logistic map, 
where successive generations in a population are related by the equation:
\begin{equation}
N_{t+1}  = r N_t [1-N_t].
 \label{May}
\end{equation}
This deceptively simple equation admits a quite complex behaviour. It has a fixed point, that is, $r N^* [1-N^*]=N^*$, for $N^* =(r-1)/r$  for $r>1$; 
but for $r>3$ the fixed point is unstable and exhibits  bifurcations and eventually chaotic behaviour for $r >3.6$ (see below). It was realised by 
Feigenbaum \cite{Feigenbaum1,Feigenbaum2} and others that these features were universal and are typical of this class of map 
(see, for instance, Ref. \cite{Gleick} for a broad description of the discovery of the chaotic behaviour). 

The possibility of a putative chaotic behaviour of the ES arises from the realisation that the logistic map might be, on a longer term, a more realistic assumption than the hypothesis Eq. (\ref{Anthropocenelinear}). This idea was developed in Ref. \cite{Bernardini2022}, where it was assumed that for a Planetary Boundary parameter between periods $n$ and $n+1$: 
\begin{equation}
h_{1(n+1)} = r\, h_{1(n)}(1 - \alpha^{-1}\,h_{1(n)}).
\label{LogisticPB}
\end{equation}
where $i=1$ was arbitrarily chosen.

This implies that the equilibrium points of the Planetary Boundaries and of the ES can show the features discussed above. Pictorially, this behaviour is depicted in Figure 2 for $\alpha=1$ \cite{Bernardini2022}. 

\vspace{0.4cm}
\begin{figure}[h!]
\centering
\includegraphics[width=0.6\columnwidth]{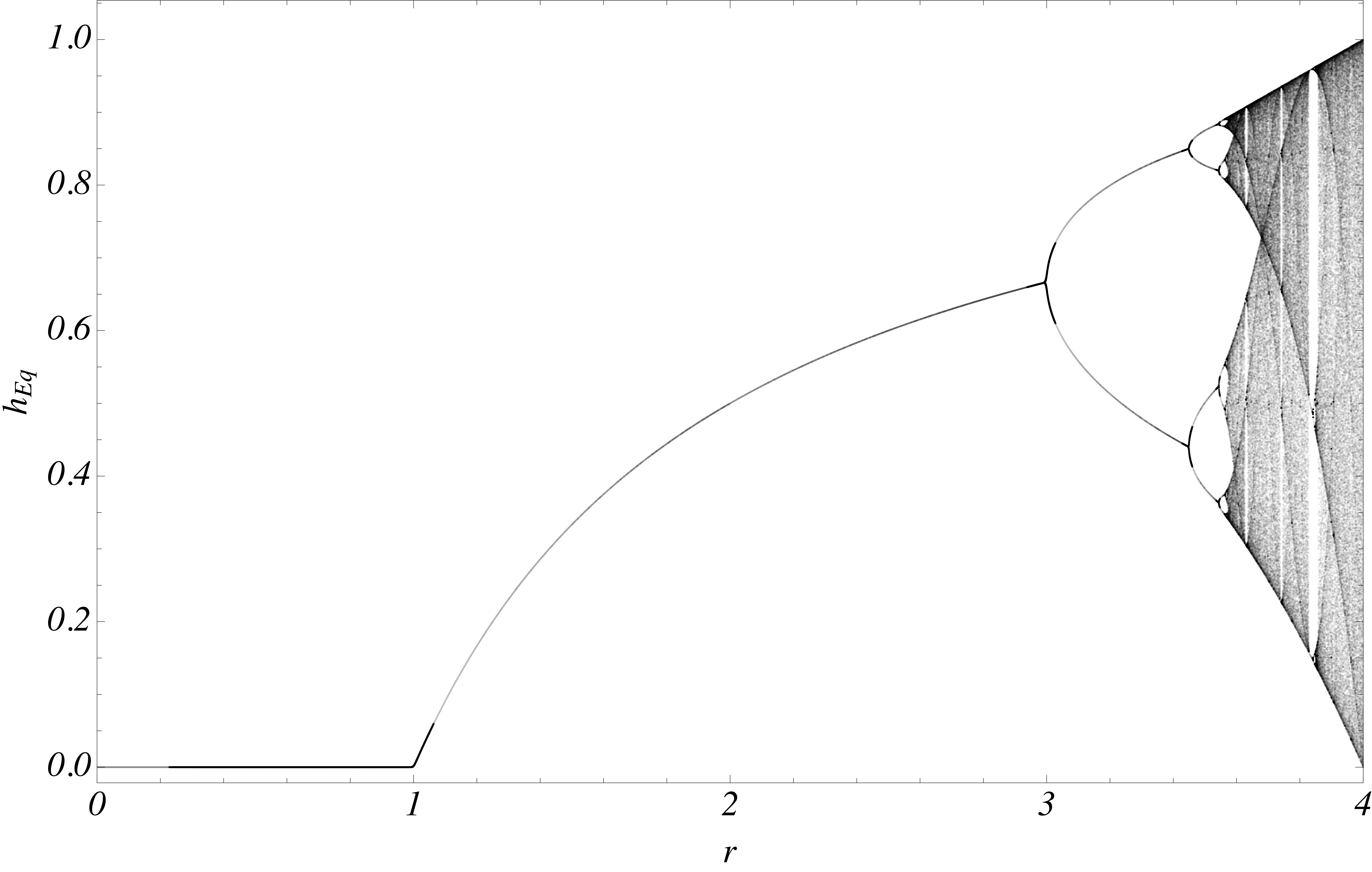}
\caption{Bifurcation diagram for the logistic map ($\alpha = 1$) \cite{Bernardini2022}.}
\label{Chao02}
\end{figure}

It follows from the Anthropocene equation (\ref{Anthropocene}), with $H\sim h_{1}$ that the equilibrium point, $\psi^{Eq}$, solution of ${dF \over dt} = 0$, 
is drastically affected by the growth rate, $r$, of $h_1$, since the equivalent map for $\psi^{Eq}\mapsto \psi^{Eq}_n$ can be obtained by
\begin{equation}\label{Novo2}
\psi^{Eq}_n \mapsto \psi^{Eq}_{n+1} = \psi^{Eq}_{n+1}(\psi^{Eq}_n), 
\end{equation}
which exhibits, as discussed in Ref.  \cite{Bernardini2022}, a similar bifurcation diagram as depicted in Figure 2.

Of course, several scenarios can be considered if the Planetary Boundaries have an hierarchy and/or depend on each other. The details are discussed in 
Ref.  \cite{Bernardini2022}. But the main point here is to emphasise that depending on the rate of change of the Planetary Boundaries, a complex behaviour pattern may emerge. 
The extreme case of the chaotic behaviour would imply, of course, that the predictability of the model to describe the behaviour of the ES would be entirely lost. 
In order to establish weather the ES is close to reach this behaviour, an analysis, based on data from the ``Great Acceleration" of the human activities and its impact on the ES  \cite{Steffen}, 
is under way \cite{Bertolami2022a}.

\section{Discussion and Conclusions}\label{Sec:Conclusion}

In this brief communication we have sketched the main features of the evolution of the ES in the Anthropocene.  
Starting with a general discussion of the fundamental discovery that the ES is an integrator of the weather 
fluctuations in which some forcing may lead to changes in the weather patterns, we have then presented an specific 
model of the ES based on the physical framework of the phase transitions. 

The proposed model [10] is fairly general and can be used to describe and predict the
behaviour of the ES as well as to be a guide for an accountancy of the human activities in
terms of the Planetary Boundaries \cite{Bertolami2020} and for the classication of rocky planets \cite{Bertolami2022b}. In what
concerns the ES, we have shown, in a series of papers, that: the impact of human activities lead the ES to new equilibrium state 
which is hotter than the Holocene state \cite{Bertolami2019}; that the interaction among Planetary Boundary terms is non negligible \cite{Bertolami2020}; 
and, more recently, to the conclusion that the evolution of the ES can exhibit complex behaviour, including a chaotic one 
if human activities behave as a logistic map \cite {Bernardini2022}. 

Of course,  the unequivocal evidence that a serious climate change crisis is under way, demands for a clear and determined global action to avert the unfolding catastrophe. 
This situation calls for stewardship measures either through global actions (see for a thorough discussion Ref. \cite{Henry}) or via the multiplication 
of local efforts to restore land and ocean ecosystems \cite{Bertolami2021a}. Most certainly these matters require a wide discussion and must be properly 
related with the broad socio-economic and political issues of our time \cite{Bertolami2021b}. Some of these issues will be discussed in this timely conference, 
in particular in the talks of Carmen Diego Gon\c calves \cite{Carmen2020}, Jo\~ao Ribeiro Mendes and others.   

\vspace{0.3cm}
\noindent 
{\bf Acknowledgements}

I would like to thank Carmen Diego Gon\c calves and Jo\~ao Ribeiro Mendes for the organisation of this very stimulant 
conference and for inviting me to be one of its speakers. I would like also
to thank my collaborators Alex Bernardini, Carlos Zarro and Frederico Francisco for the relavant comments
and suggestions.

\end{document}